\documentclass[journal]{IEEEtran}
\IEEEoverridecommandlockouts
\usepackage[T1]{fontenc}
\usepackage[latin9]{inputenc}
\usepackage{array}
\usepackage{mathtools}
\usepackage{amsmath}
\usepackage{amssymb}
\usepackage{graphicx}
\usepackage{soul}
\usepackage{xcolor}
\usepackage{subcaption}
\usepackage{tikz}
\usepackage{textcomp}

\newcommand\copyrighttext{%
  \footnotesize }

\newcommand\copyrightnotice{%
\begin{tikzpicture}[remember picture,overlay]
\node[anchor=south,yshift=10pt] at (current page.south) {\fbox{\parbox{\dimexpr\textwidth-\fboxsep-\fboxrule\relax}{\copyrighttext}}};
\end{tikzpicture}%
}
\begin{document}
\pagestyle{empty}

\title{A cloud-IoT platform for passive radio sensing: \linebreak{}
challenges and application case studies}
\author{Sanaz Kianoush,~\IEEEmembership{member,~IEEE} , Muneeba Raja,~\IEEEmembership{student member,~IEEE}, Stefano Savazzi,~\IEEEmembership{member,~IEEE},
and Stephan Sigg,~\IEEEmembership{member,~IEEE} \thanks{S. Kianoush, and S. Savazzi are with the Institute of Electronics,
Computer and Telecommunication Engineering (IEIIT) of the National
Research Council of Italy (CNR), p.zza Leonardo da Vinci, 20133 Milano,
Italy, e-mail: \{sanaz.kianoush, stefano.savazzi\}@ieiit.cnr.it.} \thanks{Muneeba Raja and Stephan Sigg are with Aalto University,Department of Communications and Networking, Finland, e-mail: \{muneeba.raja, stephan.sigg\}@aalto.fi.} \thanks{This work has been partially carried out in the framework agreement CNR-Regione Lombardia CYBER SORT project,DIT.AD008.052''.} }


\maketitle
\thispagestyle{empty}
\copyrightnotice
\begin{abstract}
We propose a platform for the integration of passive radio sensing and vision technologies into a cloud-IoT framework that performs real-time channel quality information (CQI) time series processing and analytics. 
Radio sensing and vision technologies allow to passively detect and track objects or persons by using radio waves as probe signals that encode a 2D/3D view of the environment they propagate through. 
View reconstruction from the received radio signals, or CQI, is based on real-time data processing tools, that combine multiple radio measurements from possibly heterogeneous IoT networks. 
The proposed platform is designed to efficiently store and analyze CQI time series of different types and provides formal semantics for CQI data manipulation (ontology models). 
Post-processed data can be then accessible to third parties via JSON-REST calls. Finally, the proposed system supports the reconfiguration of CQI data collection based on the respective application. 
The performance of the proposed tools are evaluated through two experimental case studies that focus on assisted living applications in a smart-space environment and on driver behavior recognition for in-car control services. 
Both studies adopt and compare different CQI manipulation models and radio devices as supported by current and future (5G) standards.  
\end{abstract}


%
\IEEEpeerreviewmaketitle

\section{Introduction}
\IEEEPARstart{P}{assive} radio sensing~\cite{wilson2011,youseff2013}, or radio vision \cite{mag2016} technology is emerging as a powerful technique for real-time processing of body-induced alterations of the radio propagation channel, enabling non-cooperative body motion recognition based on radio-frequency (RF) signatures. 
Radio vision systems for the recognition of human activities exploit the RF signals monitored within the network by internet-of-things (IoT) devices~\cite{Pervasive_Scholz_2011}.
The technology is based on the real-time processing of channel quality information (CQI) that is commonly used by wireless communication systems at the receiver-side to quantify the RF signal quality~\cite{8067693}. 
The perturbations induced by moving bodies or other objects on the electromagnetic (EM) wavefield can be measured directly from CQI and analyzed to recover a rough 2D/3D image of the environment \cite{phy}, without the need of ad-hoc sensing infrastructure nor cooperation of the monitored subjects (sensor-free)~\cite{ARTI2017}. 
The technology serves as enabler for implementing a flexible reconfiguration of sensing tasks, paving the way for a new generation of IoT systems that are made up of application-customizable sensor components, instead of application-dependent, and confront many interoperability problems arising from complex interactions between heterogeneous hardware resources.

\begin{figure}[tp]
\begin{centering}
\centering{} \includegraphics[width=\columnwidth]{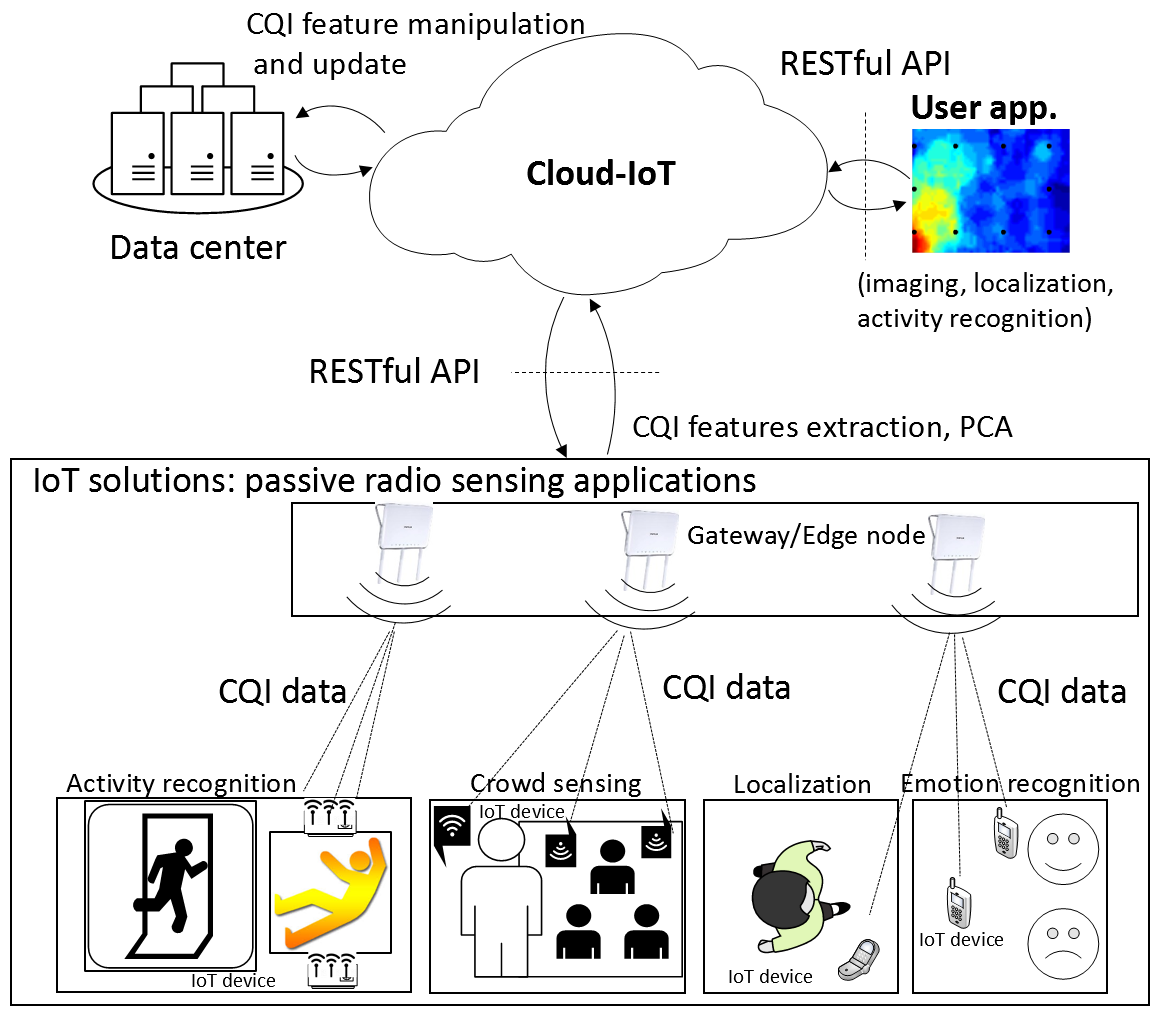} \centering{} 
\par\end{centering}
\caption{\label{cloud-IoT}Cloud-IoT for passive radio sensing including IoT field device, Gateway edge device and cloud IoT unit.}
\end{figure}

Applications of passive radio sensing are numerous and include, for instance, fall detection for ambient assisted living~\cite{wang2017wifall} and industrial workspaces \cite{IoT2016}, healthcare and elderly care applications~\cite{woznowski2017sphere}, gait monitoring, for instance in Parkinson treatment~\cite{wang2016gait}, counting of crowd in social mass gatherings~\cite{di2016trained}, as well as sleep monitoring~\cite{
liu2015tracking}. 
Due to the high frequency regimes utilized, impressive accuracy can be reached, for instance for the recognition of typing and finger movement~\cite{li2016wifinger}, lip movement~\cite{wang2016we}, or respiration sensing~\cite{wang2016human}. 
Furthermore, researchers have recently demonstrated that also attention towards surrounding objects can be captured from device-free RF sensors~\cite{feng2017mais}, as well as emotion, which is possible either via breathing and pulse~\cite{zhao2016emotion}, or, alternatively, from gestures, motion and pose~\cite{Muneeba_2017_Geospatial
}. 
\IEEEpubidadjcol

The complexity of CQI data processing/analytics required for real-time detection and tracking creates a strong need for the integration of cloud processing services \cite{cloud_complexity2017}. 
Most existing IoT platforms have been designed as isolated
vertical solutions, in which all components are tightly coupled to
the specific application context \cite{IoT2017}. On the contrary,
CQI knowledge trading among heterogeneous radio devices within the cloud
can provide a unified framework to enable smart reconfigurable sensing tasks as well as a practical solution to overcome many relevant interoperability issues of such vertical solutions. For such functionality, it is necessary
for data consumers to obtain large CQI data-sets, aggregate and isolate relevant information  from this huge bulk of clutter, and apply real-time
data processing via specific cloud services. To our best knowledge, in this paper for the first time, we propose a concise solution to extract, analyze and manipulate CQI data in real-time for the purpose of passive radio sensing that is accessible remotely through cloud resources. The proposed platform incorporates cloud-edge CQI computing services to meet different application demands: it thus provides a set of fully integrated services that allow to connect, manage and ingest CQI data from potentially globally dispersed IoT devices. 
We exploit edge caching functions~\cite{Edgecaching2017} to enable low-latency extraction of relevant ``low dimensional'' features from CQI raw data produced by the IoT radio devices. Based on such features, a common set of analytic tools is proposed for real-time inference and classification of body movements. 

As shown in Fig.~\ref{cloud-IoT}, data processing for radio sensing is done at different steps: locally inside the IoT device for raw data extraction (e.g., WiFi device/sensor, smart phones), at
edge networks (e.g., WiFi access points, gateways or sink nodes) for CQI feature computation, and globally inside the cloud section for storage and analysis. The cloud-IoT platform integrates heterogeneous CQI data models corresponding to different radio resources (e.g., 2.4GHz ZigBee, 2.4GHz-5GHz WiFi) and supports higher frequency radios \cite{win}. Also, it must apply analytics over big-sized data along with complexity consideration (typically in the order of tens of thousands of spatially distributed observations per second) collected from geographically dispersed sites (i.e. rooms, buildings, cars etc). Finally, it must enable the reconfigurable access to such massive loads of heterogeneous CQI data in the cloud, by following a sensor/actuator model as well as public/private data management profiles. 

We conduct two case studies. First, we explore the use of the cloud
platform for passive localization and tracking inside a smart-space laboratory. The CQI time-series analytics resources are accessed from a remote end-user application extracting an image of the monitored environment in real-time. The experiments are meant to validate the localization functions provided by the cloud, as well as to highlight a comparative analysis among different CQI types and features to be adopted for learning and analysis. Second, a study on in-car driver behavior detection
is carried out to experimentally validate the proposed cloud functions
in a simulated road environment. The study consists
of 40 subjects at BMW Group Research, New Technologies
and Innovation center, Germany: the system is designed
for the real-time detection of driver behavior with special focus
on distractions, possibly induced by unknown, random or external triggers. Behavior recognition resource can be accessed from vehicular network applications that might benefit from the real-time information about any risky driver activity. 

The paper is organized as follows. Sect.~\ref{sec:CQI-processing-for}
provides an overview of the radio vision technology and current state
of the art. Sect.~\ref{sec:Cloud-edge-architecture} introduces the
proposed cloud-IoT platform components. Concepts of abstraction layers
for IoT devices, edge computing (Sect.~\ref{sec:CQI-data-analytics}), as well as CQI object modeling and application layer (Sect.~\ref{sec:CQI-data-ontology}) are also addressed. Specific
case studies highlighted in Sect.~\ref{sec:Occupancy-detection-and} and \ref{sec:Activity-and-Emotion} reveal the potential of the approach in practical applications.

\section{CQI processing for passive radio sensing and vision: overview and contributions\label{sec:CQI-processing-for}}
Context and subject recognition is becoming truly pervasive, while passive radio sensing technologies provide a unique solution as capturing and extracting contextual information from radio signals exchanged among IoT devices is yet unused for sensing. 
Compared with camera-based techniques, privacy-preservation is also a second strong motivation towards the adoption of this technology. 
Despite recently proposed solutions~\cite{devicefreeSigg,Zhu2017}, the CQI-based sensing still lacks unified and standard tools to facilitate data reuse by different applications as well as the integration with other sensor platforms. 
This paper provides the basic tools to include the technology into an open cloud-IoT platform. In particular, it promotes a re-designing of CQI data processing using cloud-edge native functions and it defines conceptual data models for CQI extraction, management and real-time manipulation.


\subsection{Related work \label{subsec:Litreview}}
Recently, employing a cloud platform~\cite{Factcloud2014} for real-time managing of large scale data from physical sensors (e.g., IoT devices) is gaining attention by the research society with respect to various technologies.     
A sensor-cloud platform is proposed in \cite{Sense_cloud} to integrate cloud computing with wireless sensor networks (WSNs). The platform defines virtual sensor groups while analytic tasks can be controlled by end users to choose an appropriate service. In~\cite{IFCIoT2017}, an integrated fog computing architectural paradigm (IFCIoT) is proposed to harness the benefits of the IoT and cloud computing in a unified reconfigurable layered fog-node architecture that adapts to application requirements.

Compared with previous work, radio vision technology exploits more sophisticated CQI data processing and learning tools for recognizing body movements. 
In particular, CQI computing services can be based on heterogeneous measurements of radio signal quality collected at both physical and upper layers, including channel state information (CSI), received symbol quality~\cite{mag2016}, and network/link layer received signal strength (RSS). Measurements of CQI might include also other aggregated information including packet error rates or related metrics, and link quality information (LQI)~\cite{LQI2017}. 

The RSS is a practical metric to assess CQI at frame level, and it is commonly adopted for transmitter (TX) and  receiver (RX) link adaptation and link-layer transmission scheduling tasks. 
The RSS indicator (RSSI) can be adopted for device-free localization~\cite{Chang2017,IoT2016,ICC2016,Deak2014}. 
RSSI metrics could also apply to both device-based and device-free radio sensing tasks. 
Device-free solutions are now considered more attractive since they replace vision sensors without requiring specific radio tags~\cite{Patel2013}. 
Radio tomographic imaging (RTI)~\cite{youseff2013,yousef2014} reconstructs an image of the environment to highlight moving objects/people by tracking either the alterations of the RSS mean, the standard deviation, or both. 

The physical (PHY) layer channel response is able to discriminate multipath characteristics~\cite{shi2012activity}, and thus holds the potential to further improve the sensing accuracy~\cite{magazine2017}. 
Many recent research efforts therefore focus on using CSI and pioneering works have demonstrated sub-meter or even centimeter-level accuracy~\cite{Shi2016,CSIloc2016}. In contrast to RSS, processing of CSI can leverage multiple independent measurements at frame level and can be used to capture fast human body movements and gestures. In~\cite{Shi2016} CSI readings are derived from off-the-shelf WiFi IEEE 802.11n wireless cards which can provide fine-grained OFDM subcarrier radio measurements over multiple antennas.
In~\cite{ICC2017}, CSI information is applied at physical layer using space-frequency diversity for device-free localization. 
Human motion effects on CSI are also investigated in~\cite{Zhu2017} by aggregating the best combination of antennas to achieve more accurate and robust detection.


All above mentioned works are based on local CQI processing and are not suitable for remote operation nor designed for integration within a cloud platform. The upcoming 5G cellular system, that is envisioned to connect IoT devices, mobile personal equipment, as well as vehicular and industrial nodes, is perfectly situated to improve environmental perception from RF, by integrating these diverse radio sensors via cloud-edge aggregation, processing and management.

\subsection{Original contributions \label{subsec:maincontribution}}
Our platform provides a unified framework to connect, manage and manipulate large CQI data-sets from heterogeneous IoT devices. 
It consists of a cloud unit and Gateway devices collecting raw CQI data.
For real-time support, computing and caching runs inside the Gateway edge devices and enables segmentation and extraction of CQI features from large measurement sets. 
The cloud section processes the features received from the Gateway devices to infer a hidden process (i.e., subject motion, position or activity). 
Our framework is validated by two case studies. 
The contributions include
\begin{enumerate}
\item a cloud architecture and components for passive radio sensing, including field functionality and specification of radio devices, Gateway edge devices and platform as a service (PaaS) unit (validated by experiments);
\item cloud-edge CQI data analytics tools.
A flexible latent variable model for feature manipulation is proposed to support different sensing tasks ranging from localization to activity recognition (in-car behavior detection)
\item an ontology model \cite{semantic} for radio sensing including feature manipulation, radio link, service and control plane models;
It provides explicit semantics for CQI data and feature manipulation to simplify interactions with end-user applications and integration with existing sensor-cloud platforms and models (msm, IoT-O, oneM2M);
\item real-time cloud resources to support: i) occupancy detection and indoor localization from RSSI features, and ii) in-car behavior recognition using PHY layer CSI. 
\end{enumerate}
\section{Cloud architecture\label{sec:Cloud-edge-architecture}}

The approach described in this section lays the groundwork for
the integration of passive radio sensing and vision technologies into
a cloud framework as well as to deliver the common specifications for
new data and ontology models.
The radio sensing paradigm is integrated here within a cloud-edge platform-as-a-service (PaaS) model to enable a wide range of applications. The platform
consists of the following components, also highlighted in Fig. 2.

\textbf{Field radio devices} constitute the mote devices for the creation of multi-standard wireless sensor networks. 
The devices integrate different radio technologies to collect and measure
radio signals of different types. Various RF interfaces can be exploited for passive radio sensing, such as FM radio~\cite{wilson2011}, WiFi~\cite{
8067693}, cellular systems~\cite{singh2017smart,cellsavazzi2017}, networks of sensor nodes~\cite{devicefreeSigg}, RFID~\cite{
zou2017grfid}, short-range radar (60GHz)~\cite{
lien2016soli} or customized software radio-based systems~\cite{adib2015capturing}. 
Personal mobile devices, wearable or wrist-worn devices can also act as mobile probe RF signal generators. 
Computing and storage resources of devices are limited so that they might need to communicate via Gateway devices to reach the cloud platform (e.g. Bluetooth Low Energy (BLE) or ZigBee). 

\textbf{Gateway edge devices (GW)} provide access point and edge
computing functions for radio devices featuring diverse RF interfaces. In line with the Web-of-Things paradigm \cite{wot}, edge devices serve as RESTful-compliant web resources \cite{wot2} and they can be interconnected by a variety of communication technologies \cite{wot3}. Edge nodes also provide compute, storage, caching and communication \cite{wot-sw} between radio devices and cloud computing data center.
According to the proposed architecture, the GWs also function as CQI data collector and interact with the data center unit through the abstraction layers (Sect. \ref{sec:CQI-data-analytics}).
Each GW maintains the RF field continuously, tracks and segments alterations of the radio environment and computes CQI features (Sect. \ref{subsec:CQI-data-structuring}). 
Finally, GWs also act as over-the-air (OTA) updater to control CQI data collection based on the sensing task (Sect. \ref{subsec:Radio-device-abstractions}).

\begin{figure}[tp]
\begin{centering}
\centering{} \includegraphics[scale=0.17]{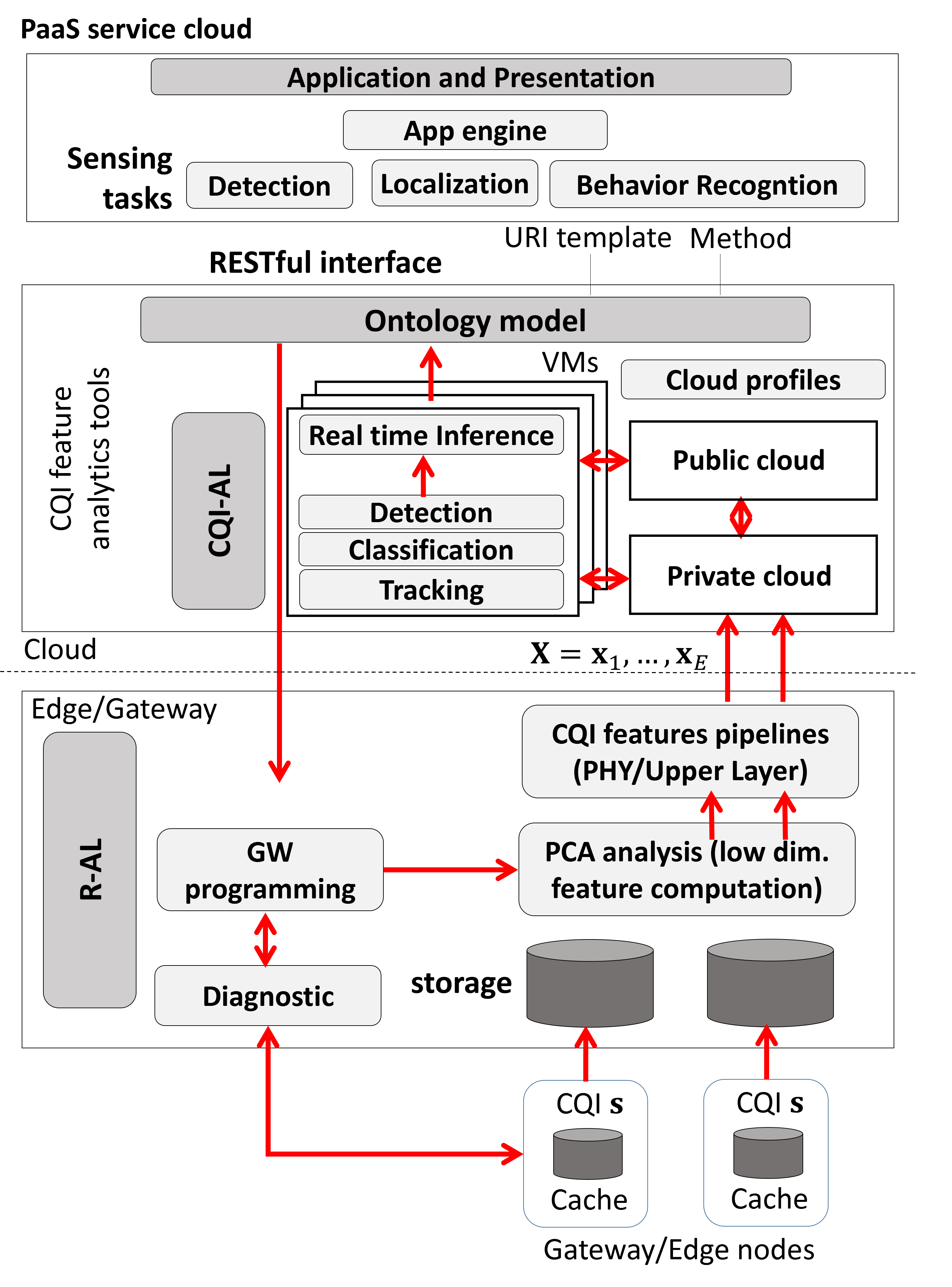} \centering{} 
\par\end{centering}
\caption{\label{cloud-edge-1}Cloud-edge architecture: R-AL for device manager,
CQI-AL and feature analysis, ontology model (OM).}
\end{figure}

Platform-as-a-service (PaaS) cloud unit to deploy and execute cloud components. PaaS models \cite{paas} are typically adopted to provide a unified means of connecting different applications to backend cloud storage and computing, while also delivering common IoT functions such as end-user management and authentication, push notifications, and other features in accordance with the specific use case.
The cloud unit is here designed to enable real-time manipulation of heterogeneous CQI
feature streams. The unit also provides an intermediate, open layer
between third party applications and underlying CQI manipulation resources (based on JSON-REST calls \cite{wot-sw}), keeping all the complexity and functionalities necessary to implement/manage the latter. As depicted in Fig. \ref{cloud-edge-1}, the cloud section consists of a private and a public profile for data management. In particular, the private cloud~\cite{personalCloud} facilitates the users to define the features they don't want to share in the public profile. 
Further, it can generate personalized feedback.
Both private and public sections consist of three layers: 
\begin{itemize}
\item[i)] the radio abstraction layer (R-AL) provides a southbound interface to individual GW devices producing CQI features from raw RF data and controlled by the cloud;
\item[ii)] the CQI data layer provides a common application programming interface (API) for real-time feature manipulation; 
\item[iii)] the ontology layer defines the CQI manipulation and task configuration resources available to applications.
\end{itemize}
These components are described in the following section. 

\section{Cloud-edge CQI data analytics tools\label{sec:CQI-data-analytics}}
\begin{figure}[tp]
\begin{centering}
\centering{} \includegraphics[scale=0.135]{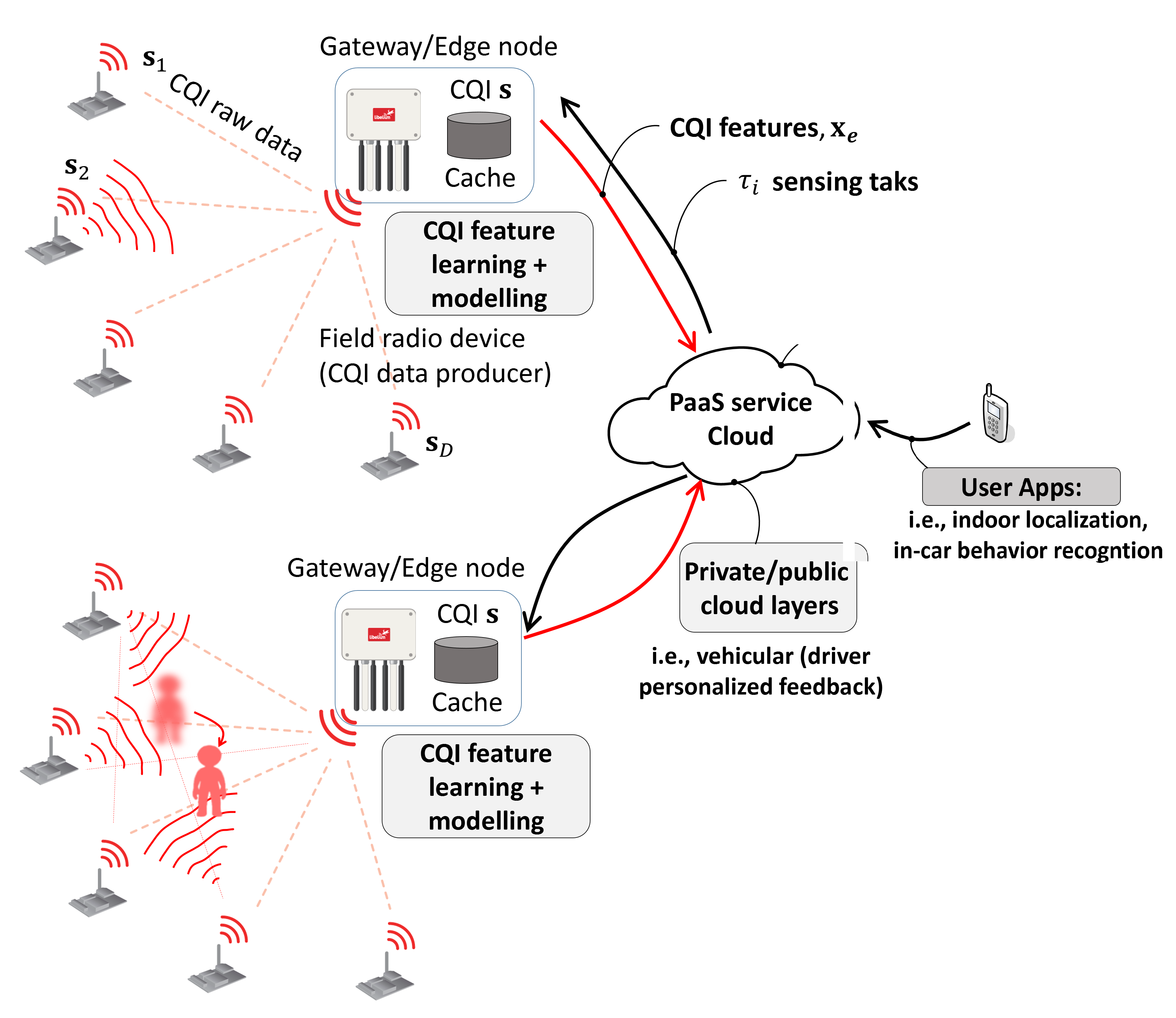} \centering{} 
\par\end{centering}
\caption{\label{cloud-edge}Cloud-edge platform for passive radio sensing}
\end{figure}

In this section, we introduce the CQI data analytics and abstraction tools that decouple the application-dependent passive radio sensing tasks from the pre-processing of raw CQI data. 
As illustrated in Fig.~\ref{cloud-edge}, we consider a cloud-IoT architecture with $E$ GW nodes which can serve a group of $D$
field devices deployed in the detection area. The GWs are connected
to the cloud by means of a transport network and are equipped with local
caches for raw CQI data extraction (Sect.~\ref{subsec:Radio-device-abstractions}), pre-processing  and storage of trained/learned features. 
In addition, an actuation link can be used by the cloud section to activate new sensing tasks on a specific GW: such tasks are here labeled as $\tau_{i}$. 

Edge computing functions are exploited inside the GW node to enable the segmentation, the extraction and the delivery of CQI features (Sect.~\ref{subsec:CQI-data-structuring}). 
Features serve as inputs for task-dependent classification problems that are tackled by the cloud section. For each task $\tau_{i}$, a discrete (latent) process $z_{i}$ that is hidden in the received feature series is identified through classification. 
Latent process values, i.e., $z_{i}=k$, indicate task-relevant information to be detected, such as subject presence/absence, location or behavior (Sect.~\ref{subsec:CQI-feature-manipulation}). 
Finally, computational overhead is assessed for CQI manipulation and REST interface management (Sect.~\ref{subsec:computation_overhead}). 

\subsection{Radio device abstraction and CQI data types \label{subsec:Radio-device-abstractions}}
The radio abstraction layer (R-AL) interacts with radio hardware to program specific radio functions (if applicable) or to manage CQI data. 
It runs inside the edge node performing data collection and device programming. 
The former is based on a categorization of the CQI measurements (CQI\_TYPE). 

\begin{itemize}
\item[i)] \emph{physical (PHY) layer} channel quality information at baseband symbol level includes channel state information (CSI), and received symbol quality~\cite{mag2016};
\item[ii)] \emph{upper layer (UP)} network/link-layer received signal
strength (RSS)~\cite{wilson2011} or other aggregated
channel quality information includes packet error rates or related
metrics, and link quality information (LQI);
\item[iii)] \emph{raw signals (IQ)} include raw features such
as micro-Doppler measurements, dynamic phase shifts and IQ channel
envelope~\cite{iot22}.
\end{itemize}

The R-AL implements specific over-the-air (OTA) device programming functions (OTA\_FUNCTION) that can be used to modify the CQI data collection process, and, in turn, the sensing task itself. 
Programming and updates are controlled by the edge GW node and are in the form of low-level firmware or upgrades. 
Gateway devices forward the OTA commands received from the cloud section and thus act as OTA updater, while field devices act as OTA update receivers. Each code module implements a specific OTA profile that maps onto a target sensing task. Table \ref{tab} lists main supported profiles.

\begin{table*}
\centering{}%
\begin{tabular}{|l|l|}
\hline 
\textbf{OTA\_FUNCTION}  & \textbf{Reconfiguration description}\tabularnewline
\hline 
\hline 
One-hop neighborhood  & number of active links to cover a detection area, or device-to-device
neighborhood \tabularnewline
\hline 
Frequency and bandwidth  & operating carrier frequency/channel, (subcarriers for OFDM radio interfaces)\tabularnewline
\hline 
Transmission duty cycle  & time interval between two consecutive radio transmissions that rules
the RF signal emission rate \tabularnewline
\hline 
CQI type  & reconfiguration (where applicable) of the CQI type - PHY, UP, IQ\tabularnewline
\hline 
CQI sampling  & reconfiguration of the CQI sampling (during debugging) \tabularnewline
\hline 
\end{tabular}\medskip{}
 \caption{Selected radio functions for OTA programming profiles.}
\label{tab} 
\end{table*}

\subsection{CQI data abstraction and features \label{subsec:CQI-data-structuring}}
Device-free radio sensing techniques typically generate and process
huge amounts of CQI data that require ad-hoc management tools not supported by traditional databases. In-fact, beside storage, the database service provider should take the responsibility for CQI data processing and time series analysis that are more complex compared with conventional sensor networks applications. For example, to support passive subject localization and radio imaging functions (Sect. \ref{sec:Occupancy-detection-and}), about $200$ wireless links, or more \cite{IoT2016}, are typically monitored per room/building. 
For each link, CQI must be measured and processed in real-time on every tens of milliseconds, to track body motion. 
Activity recognition exploiting WiFi signals (i.e., inside cars, Sect. \ref{sec:Activity-and-Emotion}) typically requires CQI processing at PHY layer, or CSI, with symbol time of hundreds of micro-seconds up to few milliseconds, and aggregates multiple time series each representing individual OFDM subcarriers and/or WiFi antennas \cite{magazine2017}.  

CQI analysis for passive subject recognition should be based on some form of decentralized processing: this is implemented here by a cloud-edge computing framework. As depicted in Fig.~\ref{cloud-edge}, the proposed approach lets each edge node $e$ pre-manipulate the raw CQI data collected
from the field devices to obtain corresponding \emph{CQI features}
represented as the vector $\mathbf{x}_{d,e}$. 
These are low-dimensional representations of CQI time series $\mathbf{s}_{d}$ obtained from field device $d$, namely by transformation $\mathbf{x}_{d,e}=\mathrm{F}_{e}(\mathbf{s}_{d})$. 

CQI data $\mathbf{s}_{d}$ might cover different physical wireless links
$\ell_{j}\in\mathcal{L}$ (or antennas) and carriers/sub-channels $\text{ }f_{m}\in\mathcal{F}$. 
The raw data vector is modeled as:
\begin{equation}
\mathbf{s}_{d}=\left\{ s_{v}=s_{v,0}+\triangle s_{v}\right\} _{v=\left[f_{m},\ell_{j},t\right]},\label{raw_data}
\end{equation}
each observed sample $s_{v}=s_{f_{m},\ell_{j},t}=s_{v,0}+\triangle s_{v}$
is extracted using the R-AL interface and is a function of the time
instant $t\in[t,T-t]$, the frequency $f_{m}$ and the link $\ell_{j}$,
here combined in the index $v$. 
The terms $\triangle s_{v}$ capture the corresponding time-varying alterations of the channel response, compared to background $s_{v,0}$, as induced by body movements, activities or changing scene.

For an assigned task $\tau_{i}$, the CQI features $\mathbf{x}_{d,e}$ are computed by implementing a probabilistic PCA analysis~\cite{PCAtipping} (see also ~\cite{PCA2002} for a review) of the input CQI data~(\ref{raw_data}). 
The goal is to reduce the dimension of the CQI data-set, while still preserving the information about task-relevant alterations induced by body movements, to enable remote manipulation. 
The CQI features corresponding to the input data vector $\mathbf{s}_{d}=\left\{ s_{v}\right\} _{v=1}^{V}$, of size $V$, are obtained as
\begin{equation}
\mathbf{x}_{d,e}=\mathbf{U_{\mathit{P}}^{\mathit{T}}}\mathbf{s}_{d},\label{eq:feat}
\end{equation}
with subspace $\mathbf{U}_{\mathit{P}}=[\mathbf{u}_{1},\mathbf{u}_{2},...,\mathbf{u}_{P}]^{T}$.
The features describe the first $P<V$ principal components of the sample
covariance matrix $\mathbf{C}_{\tau_{i}}$ 
\begin{equation}
\mathbf{C}_{\tau_{i}}=\sum_{z_{i}=k}\mathrm{Pr}(z_{i}=k)\frac{1}{D}\left(\sum_{d=1}^{D}\mathbf{\overline{s}}_{d}(k)\mathbf{\overline{s}}_{d}^{T}(k)\right).\label{eq:mar-1-1}
\end{equation}
Covariance $\mathbf{C}{}_{\tau_{i}}$ depends on the sensing task
$\tau_{i}$ and its corresponding latent process values $z_{i}=k$ with prior probability $\mathrm{Pr}(z_{i}=k)$\footnote{Uniform prior can be used when no prior knowledge is available.}.
It is computed by marginalizing out the sample covariances $\sum_{d=1}^{D}\mathbf{\overline{s}}_{d}(z_{i}=k|\tau_{i})\mathbf{\overline{s}}_{d}^{T}(z_{i}=k|\tau_{i})$
obtained from training data examples $\mathbf{\overline{s}}_{d}(k)=\mathbf{\overline{s}}_{d}(z_{i}=k|\tau_{i})$.
Training data $\mathbf{\overline{s}}_{d}(k)$ provides the ground-truth
for each considered latent process value $z_{i}=k$~\cite{ICC2016,magazine2017}. 

The selection of the principal components depends on the specific sensing task $\tau_{i}$ (see Sect. \ref{sec:Occupancy-detection-and} and \ref{sec:Activity-and-Emotion}
for case studies). 
Notice that, before manipulation, raw data can be pre-processed at edge nodes for denoising and interpolation~\cite{mag2016} as well as for windowing and segmentation. 
In addition to PCA, other processing might be required, based on the sensing tasks to describe the CQI body-induced footprints. An example is
given in Sect. \ref{sectionBehaviorRecognition}.

\subsection{CQI feature manipulation and data management\label{subsec:CQI-feature-manipulation}}
Feature manipulation and related management tasks are carried out by the cloud section. 
For a sensing task $\tau_{i}$, the objective is to infer a discrete latent process $z_{i}$ which takes values $k\in\left\{ 1,...,K_{i}\right\}$. 
For positioning and detection, the latent process values $z_{i}=k$ correspond to $K_{i}$ monitored locations selected by the user. 
In contrast, for the considered activity recognition study of Sect. \ref{sec:Activity-and-Emotion}, these refer to distracted driver behaviors.

For the task $\tau_{i}$, we assume that edge GW node $e$
is producing feature vectors $\mathbf{x}_{e}=[\mathbf{x}_{1,.e}^{T},...,\mathbf{x}_{D,e}^{T}]^{T}$
consisting of $D$ components $\mathbf{x}_{d,e}$, each corresponding to a connected field device. Features to be computed by
the GW can be configured on-demand and are based on PCA
analysis (equation~(\ref{eq:feat})). 
The latent process value is inferred based on a metric $\mathrm{M}(\mathbf{X}|\tau_{i})$ applied to the CQI features $\mathbf{X}=[\mathbf{x}_{1},...,\mathbf{x}_{E}]$
from the $E$ edge nodes configured for the same task. 
It consists of the mixture components 
\begin{equation}
\mathrm{M}(\mathbf{X}|\tau_{i})=\sum_{k=1}^{K_{i}}\alpha_{k}\sum_{e=1}^{E}\mathrm{G}^{k}(\mathbf{x}_{e}|\tau_{i}).\label{eq:ugm-2}
\end{equation}
In particular, probability terms $\alpha_{k}=\mathrm{Pr}(z_{i}=k)\in[0,1]$
represent the prior knowledge on latent process $z_{i}$, with $\sum_{k=1}^{K_{i}}\alpha_{k}=1$.
Mixture component $\mathrm{G}^{k}(\mathbf{x}_{e}|\tau_{i})$ describes the similarity between the observed features $\mathbf{x}_{e}$ and the expected ones according to the latent process value $k$.
Such metric depends on the sensing task and is considered in the case
studies of Sect. \ref{sec:Occupancy-detection-and}-\ref{sec:Activity-and-Emotion}.

The real-time inference of the latent process extracts the most likely
value of the hidden variable $z_{i}$ as 
\begin{equation}
\hat{z_{i}}=\mathrm{\arg\max_{\mathit{k}\in\left\{ \mathit{\mathrm{1},...,K_{i}}\right\} }}\mathrm{Pr}(z_{i}=k|\mathbf{X};\tau_{i})\label{classification}
\end{equation}
with probability
\begin{equation}
\mathrm{Pr}(z_{i}=k|\mathbf{X};\tau_{i})=\frac{\alpha_{k}\sum_{e=1}^{E}\mathrm{G}^{k}(\mathbf{x}_{e}|\tau_{i})}{\sum_{h=1}^{K_{i}}\alpha_{h}\sum_{e=1}^{E}\mathrm{G}^{h}(\mathbf{x}_{e}|\tau_{i})},\label{eq:ugm-1}
\end{equation}
for the input evidence $\mathbf{X}=[\mathbf{x}_{1},...,\mathbf{x}_{E}]$.
\subsection{Computational overhead assessment \label{subsec:computation_overhead}}
The edge and the cloud sections, respectively introduce a computational overhead for CQI manipulation and REST interface management. In particular, the edge node is here implemented on a low-power single board embedded computer (dual core CPU, pico-itx form factor).  Beside controlling the sensing, and supervising the initial training stage, most significant overhead is due to the real-time computation of the PCA components to extract features in equation (\ref{eq:feat}). Using the Bachmann-Landau notation, complexity scales with $\mathcal{O}(VP)$, with $P$ being the number of components and $V$ the size of input CQI data. Notice that input data $V$ depends on the number of monitored wireless links for the considered sensing task (antennas or sub-carriers, for multi-carrier systems). Run-time complexity of the recognition tools implemented inside the cloud section depends on the sensing task $\tau_{i}$. Mixture components $\mathrm{G}^{k}(\mathbf{x}_{e}|\tau_{i})$ measure similarities with trained features while their computational overhead scales with $\mathcal{O}(P^{2})$ or  $\mathcal{O}(VP)$ when reconstruction of the CQI data from the input features is required. Complexity of the classification stage on $K_{i}$ latent process values scales linearly as $\mathcal{O}(K_{i})$.

Finally, as detailed in the following section, we chose to adopt JSON (JavaScript Object Notation) for encoding the features and sending through the REST interface. JSON parsing adds additional computation and size overhead as the result of object serialization/deserialization stages, however many open source libraries can be found \cite{json} to optimize the performance. JSON encoding/decoding overhead is assessed experimentally in Sect. \ref{sec:Occupancy-detection-and}.

\begin{figure*}
\begin{centering}
\centering{} \includegraphics[scale=0.133]{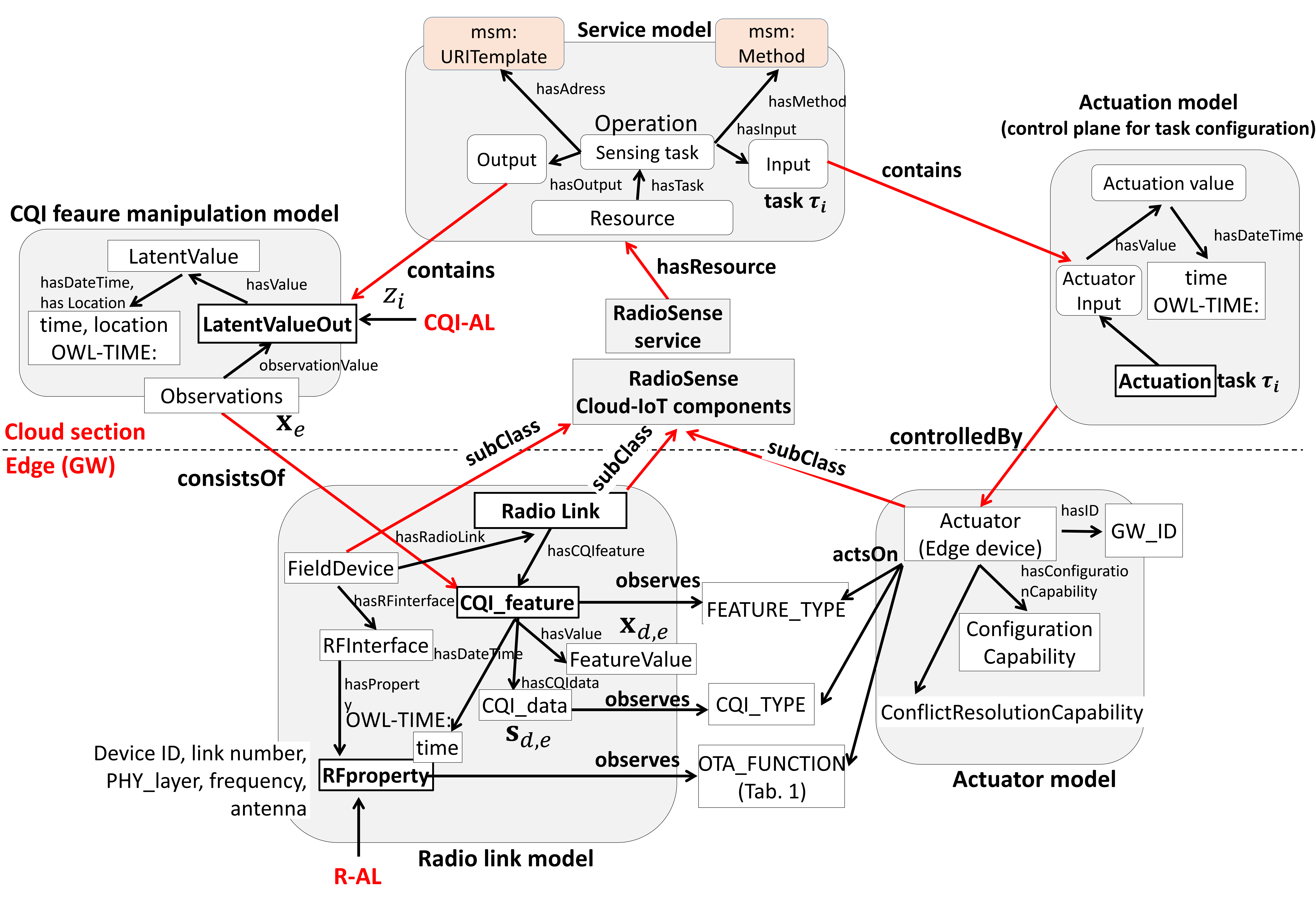} \centering{} 
\par\end{centering}
\caption{\label{ontology} Ontology model for radio sensing based on IoT-O and msm}
\end{figure*}

\section{Ontology modeling and application layer\label{sec:CQI-data-ontology}}
An ontology model is proposed to provide
formal explicit semantics for CQI feature manipulation and management
so as to facilitate reuse by different applications as well as integration
with conventional sensor-based cloud platforms \cite{m2m_semantic}. 
Object models (OMs) are defined to enable end-user applications to interact with the feature analysis and manipulation tools (Sect. \ref{sec:CQI-data-analytics}), as well as to communicate with the end user, subject to the provided authorization and permission controls. 
Figure~\ref{ontology} depicts the ontology composed of five components: the radio link abstraction from which the feature vectors are extracted,
the CQI feature manipulation, the actuator (here represented by the
edge GW device), the actuation for sensing reconfiguration, and the
service model.

Compared with standard designs, some OMs needs to be redefined for the passive radio sensing problem. The semantic models for CQI manipulation and services let the application layer efficiently process CQI features based on the selected
sensing tasks. We select the minimal service model
(msm) ontology\footnote{http://www.onem2m.org/technical/onem2m-ontologies} to describe services since it provides
a common vocabulary based on existing web standards.
Access to CQI features, manipulation
and configuration of tasks are represented as web resources to provide
the defined functions to third party users. The radio link and actuator/actuation
models allow the cloud section to act on CQI data with the purpose of triggering a change of the sensing task (when
needed by the assigned operation/task or application). These specific models comply with the IoT-O~\cite{semantic}, now integrated within the oneM2M standardization~\cite{m2m_semantic}. 

\subsection{CQI feature manipulation model}
The semantic model for CQI manipulation represents both streaming data
(CQI features) and the associated metadata information: time, location, edge node,
RF interface parameters (such as link $\ell_{j}$, and carrier $f_{m}$). 
The cloud offers a web service used by the edge nodes for streaming CQI features obtained from radio link.
Pushing of new features $\mathbf{x}_{e}$ is implemented by sending
a request to the address ``POST \textbackslash{}api\textbackslash{}post\_CQIfeatures\textquotedblright .
Given that parsing large-sized CQI data using XML files can significantly impair performance, especially under real-time constraints, JSON (JavaScript Object Notation \cite{json}) is adopted for structuring
the features and the corresponding metadata for each device. Manipulation directly
interfaces with the tools obtained from the CQI-AL (Sect. \ref{subsec:CQI-feature-manipulation}):
it thus implements feature processing for assigned sensing tasks
$\tau_{i}$ and outputs the real-time latent variable decisions $\hat{z_{i}}$ (\ref{classification}) or probabilities (\ref{eq:ugm-1}).
Third-party apps periodically retrieve such results in the form of JSON objects via the WebSocket protocol. 
The Socket server responds to GET requests on ``\textbackslash{}api\textbackslash{}SENSING\_TASK\textbackslash{}get\_LatentValues?GW=GW\_ID\textquotedblright{} for a specific operation (SENSING\_TASK) $\tau_{i}$ and edge node
$e$, with identification 'GW\_ID'. 
The retrieval of a latent process can be also available in the form of pull requests (for near real-time applications) and JSON feeds. 
Finally, the latent variable results are interpreted by the remote application based on user needs (i.e. location or activity recognition).

\subsection{Radio link model }
Radio sensing turns radio links, commonly adopted for machine-type communication, into opportunistic sensors to extract an understanding of the physical space, thereby replacing or augmenting ad-hoc metering devices. Therefore, beside CQI feature manipulation, the ontology model must include an explicit semantic for the radio link, acting as CQI data producer. 
In Figure~\ref{ontology} the link model consists of multiple CQI feature objects and it is associated with a field device that, in turn, might have multiple active radio links. 
The CQI features are extracted from CQI data sets according to the actuator
commands (FEATURE\_TYPE and CQI\_TYPE). 

The RF interface is an abstraction of the physical radio interface and has
RF properties (detailed in Table I), obtained from the R-AL. 
These can be configured by the GW edge device on demand (OTA\_FUNCTION).

\subsection{Service model and sensing tasks}
The model complies with existing oneM2M based ontologies\footnote{http://www.onem2m.org/ontology/Base\_Ontology/oneM2M\_Base\_Ontology-V\_3\_3\_0.owl}.
It defines sensing tasks that can be instantiated by third-party apps
and combined to build upper-layer services. The implemented platform
currently supports three tasks, namely detection, localization and activity recognition. 
The cloud section provides resources to activate a new task on a specific edge GW node. 
This is done by sending requests to the address ``\textbackslash{}api\textbackslash{}SENSING\_TASK\textbackslash{}start\_task?GW=GW\_ID\textquotedblright.
Such activation of a new task then triggers the control plane.

\subsection{Control plane: actuator and actuation models}
In radio sensing, a change, or activation, of a new task can be done
in real-time and it requires an OTA modification of the CQI data collection
and CQI feature manipulation processes. Based on the selected task,
data collection can be changed by the cloud through a JSON request-response
stage: the GW node periodically monitors the resource ``\textbackslash{}api\textbackslash{}SENSING\_TASK\textbackslash{}CQI\_feature\textquotedblright{}
that contains the list of active tasks $\tau_{i}$ for each installed
GW, as well as features to compute. 
Figure~\ref{ontology} illustrates the corresponding ontology model that is made of the actuator,
actuation (or reconfiguration), and sensing task information. The
actuator model represents the information about the available radio
links and actuating capabilities, including CQI features, types and
other OTA programmable functions. The actuation model describes the
configuration commands that are necessary to implement the assigned
task. For tasks that require a device firmware change, the edge GW is acting as the OTA updater.

\section{Occupancy detection and Localization\label{sec:Occupancy-detection-and}}
\label{OD_loc} We consider localization and occupancy detection case studies to evaluate real-time feature manipulation tools described above. 
We exploit target detection using both RSSI and CSI measurements and collected from WiFi and ZigBee (IEEE 802.15.4) radio devices. 

The WiFi devices (Fig.~\ref{plotCQI}) are equipped with 3 antennas and work in 5.32GHz as orthogonal-frequency division multiplexing (OFDM) radio interfaces. 
The utilized Intel Wireless Link 5300 (IWL5300)\footnote{Linux 802.11n CSI tool, https://dhalperi.github.io/linux-80211n-csitool/} reports CSI data from 30 sub-carriers as well as RSSI data from each of the $3$ antennas. The IEEE 802.15.4 nodes (Fig.~\ref{cloud_iot_loc}) measure RSSI, over the 2.4GHz band every $60$ms (corresponding to the Physical Protocol Data Unit - PPDU - super frame length). 

The GW edge node provides access point functions and communicates with the remote cloud section. 
It collects raw CQI data (RSSI and CSI) and stores them in the local cache. 
For each supported task $\tau_{\mathit{i}}$ with associated latent variable set $z_{i}$, it is assumed that it has a backlog of training data-sets, namely $\mathbf{\overline{s}}_{d}(z_{i}=k|\tau_{i})$ $\forall k,\tau_{i}$, as ``true examples'' obtained in the same environment during a calibration stage \cite{mag2016}.
These samples reveal the alterations of CQI data corresponding to true subject locations. 
Note that training data is used locally by the edge node for CQI feature computation, and is not posted on towards the cloud back-end. 

\subsection{Feature extraction and analysis}
\begin{figure}
	\begin{centering}
		\includegraphics[scale=0.55]{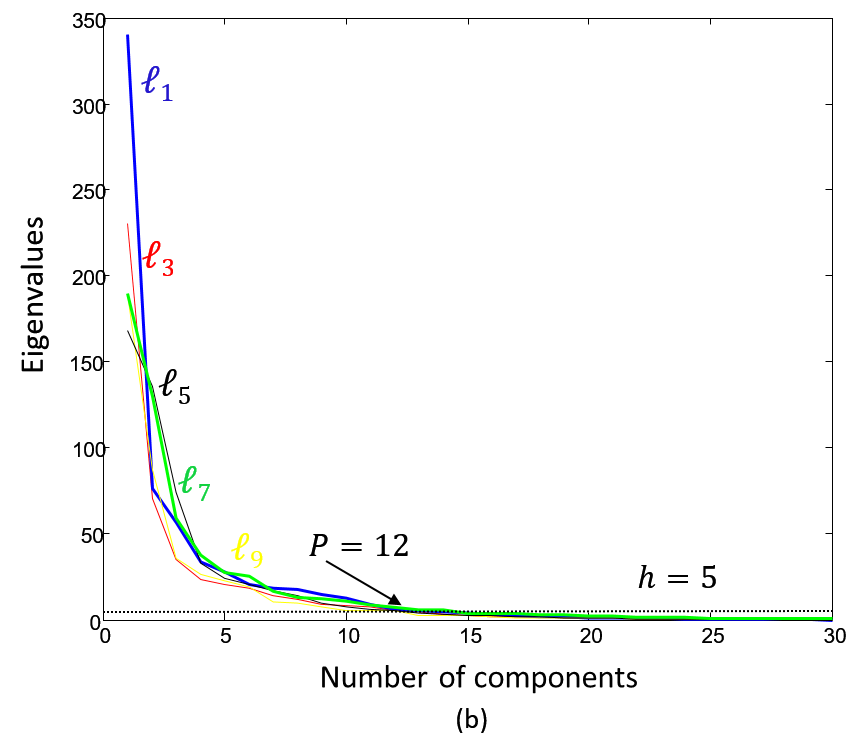}
		\par\end{centering}
	\caption{\label{PCA} Example of PCA analysis at training phase done inside one edge node for detection of 2 targets (see layout in Fig. \ref{plotCQI} (c)) using WiFi devices and the corresponding links $\{\ell_1,\ell_3, \ell_5,\ell_7, \ell_9\}$. $\mathit{P}=12$ components are selected to meet the condition on eigenvalues $\geq h$.}
\end{figure}

PCA analysis is done inside the edge node by taking the largest $P$ eigenvalues $\mathbf{U}_{\mathit{P}}$ in equation~(\ref{eq:feat}) of the marginalized covariance
matrix $\mathbf{C}_{\tau_{i}}$. For the passive localization problem,
marginalization (equation \ref{eq:mar-1-1}) is applied to capture the most
significant signal changes due to the target located inside the detection
area. In the example of Fig. \ref{PCA}, PCA is carried out
on CSI data obtained from WiFi devices. 
It selects, for each time instant, the components that are more
sensitive to the target presence. 
Analysis is carried out at training phase, based on the measurements collected from different links $\{ \ell_1, \ell_3, \ell_5, \ell_7,\ell_9\}$. 
It shows that $40\%$ of the subchannels ($\mathit{P}=12$ out of $V=30$ available subcarriers for each OFDM symbol) are sufficient to capture the most significant alterations of CQI as induced by target presence. Selection is made by setting a threshold value on the
eigenvalue (i.e., here threshold value is $h=5$).

\begin{figure}
	\begin{centering}
		\includegraphics[scale=0.19]{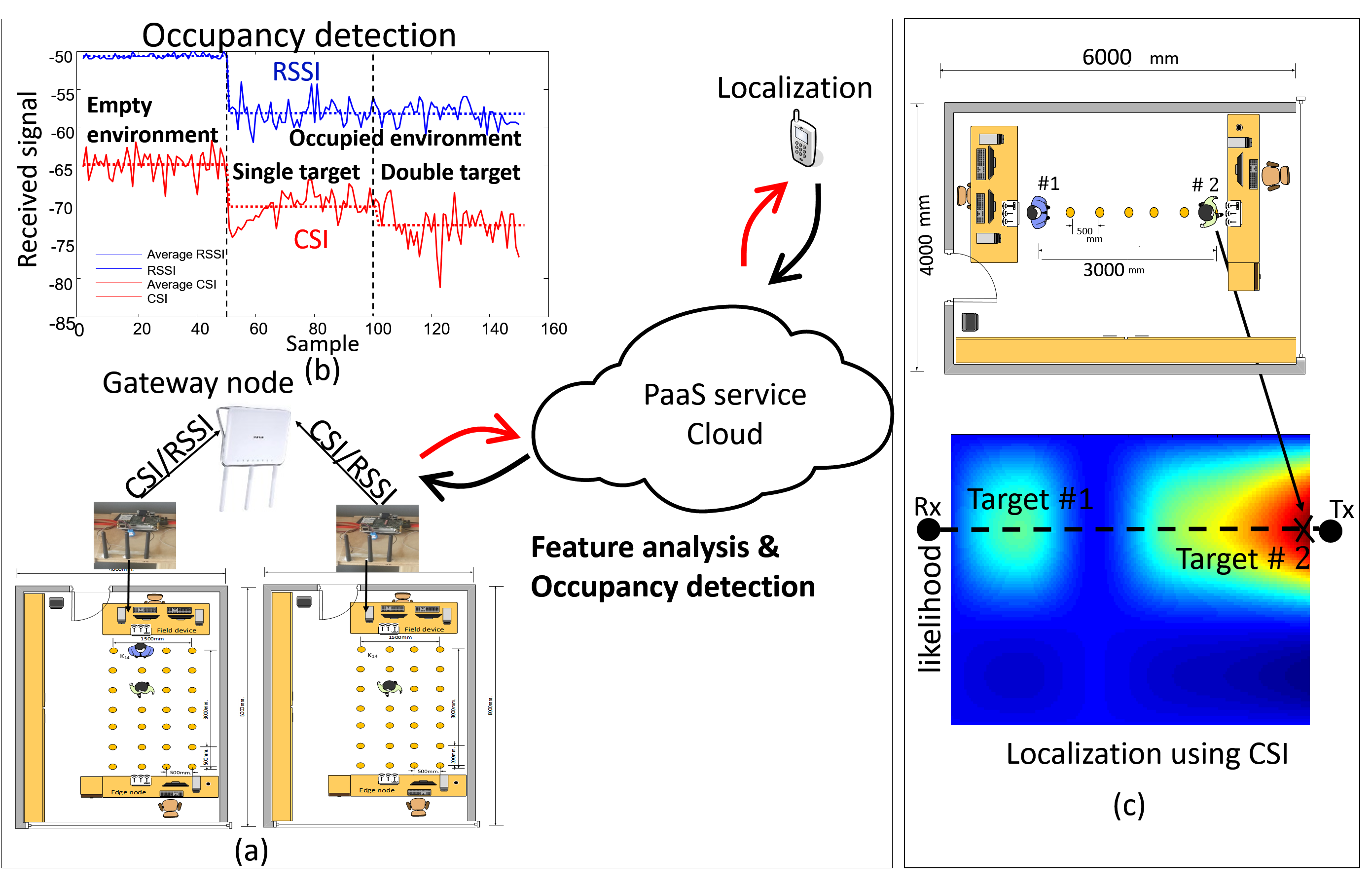} 
		\par\end{centering}
	\caption{\label{plotCQI} CSI and RSSI based occupancy detection and localization: a) layout with single and double targets, b) received signal for both RSSI and CSI types for empty and occupied environment, c) localization (top) and corresponding imaging app. (bottom).}
\end{figure}

\begin{figure*}
\begin{centering}
\includegraphics[scale=0.19]{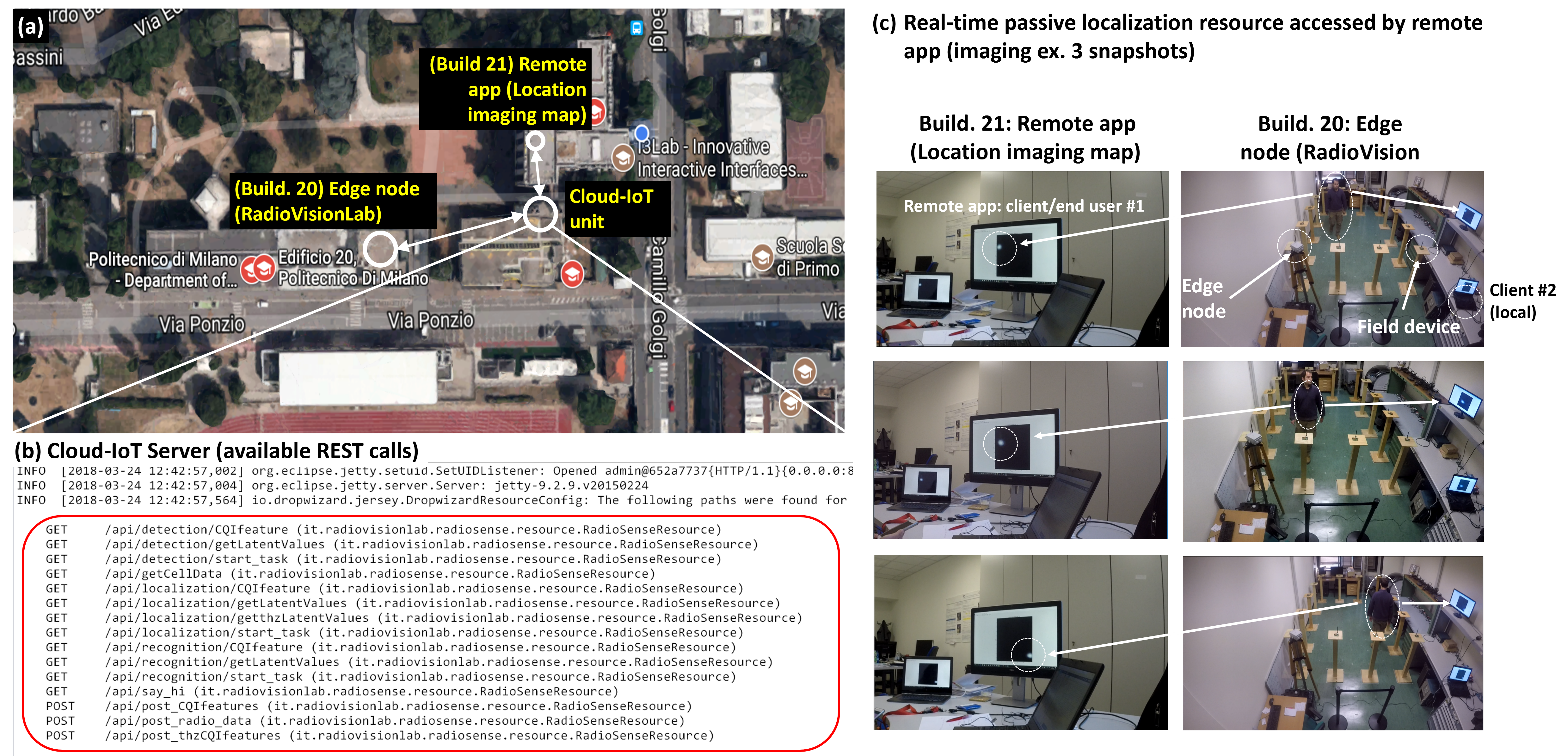} 
\par\end{centering}
\caption{\label{cloud_iot_loc} Real-time cloud-IoT based localization: a) map of the demo setting (CNR-IEIIT Milano Site, c/o Politecnico di Milano, DEIB department, buildings 20 and 21); b) list of available REST calls and resources inside the cloud server; c) ground-truth video snapshots from  experiments in building 20 and imaging applications run simultaneously in building~21.}
\end{figure*}

For feature manipulation, we adopt the mixture model (\ref{eq:ugm-2})
where latent values $z_{i}$ are interpreted as detectable
positions (see Fig. \ref{plotCQI}). 
The cloud receives the features $\mathbf{x}_{e}$ from the corresponding edge GW node.
Then it performs the reconstruction of the original CQI data $\mathbf{\hat{s}}=\mathbf{U}_{\mathit{P}}\mathbf{x}_{e}$ before PCA.
Note that the subspace projection matrix $\mathbf{U}_{\mathit{P}}$,
obtained by the GW node during training, must be passed to the cloud
in advance i.e., during task configuration. 
The log-likelihood function on reconstructed CQI samples
$\mathbf{\hat{s}}$
is used as similarity metric
\begin{equation}
\mathrm{G}^{k}(\mathbf{x}_{e}|\tau_{i})=\log\left[\mathrm{Pr}\left(\mathbf{\hat{s}}=\mathbf{U}_{\mathit{P}}\mathbf{x}_{e}|z_{i}=k\right)\right].\label{eq:likelihood}
\end{equation}
Finally, the real time inference of subjects locations (\ref{classification}) conforms with Maximum Likelihood (ML) and uses a uniform prior, $\alpha_{k}=\mathrm{Pr}(z_{i}=k|)=\frac{1}{K_{i}}$.
 

\begin{table}[!t]
	\global\long\def\arraystretch{1.3}
	\caption{Occupancy detection performance}
	\label{Detresult} \centering %
	\begin{tabular}{|c|c|c|c|c|}
		\hline 
		& Sensitivity  & FPR  & Accuracy  & Specificity\tabularnewline
		\hline 
		CSI  & 0.99  & 0.001  & 0.99  & 0.99\tabularnewline
		\hline 
		RSSI  & 0.83  & 0  & 0.84  & 1\tabularnewline
		\hline 
	\end{tabular}
\end{table}

\subsection{RSSI vs CSI analytics}    
The detector of body presence classifies the environment as empty or occupied
(by one or more subjects) via $z_{i}=\left\{ 0,1\right\} $. 
In Fig.~\ref{plotCQI} we consider 2 subjects moving inside a laboratory space. Detection is obtained by WiFi signal inspection. Fig.~\ref{plotCQI} (b) shows received signal sensitivity to the single/double target presence using both RSSI and CSI measurements, for empty and occupied environments. 
Table~\ref{Detresult} shows occupancy detection results in terms of sensitivity, false positive rate (FPR), accuracy and specificity~\cite{IoT2016}.
With respect to RSSI, CSI results improve sensitivity and accuracy
to $99\%$.

When subjects are detected, a device-free localization procedure is triggered by the cloud section on the specific edge node, using the resource ``\textbackslash{}api\textbackslash{}localization\textbackslash{}start\_task\textquotedblright.
As highlighted in the previous sections, CQI features might be based on RSSI or CSI samples: we thus perform a comparative analysis of their performance using a single link. 
In Fig~\ref{plotCQI} (c) the ML criterion (equation~(\ref{eq:likelihood})) is applied for positioning (equation~(\ref{classification})) and considers 2 targets moving in the surrounding of a WiFi link. The
imaging map (obtained from CSI data) in the same figure can be used
to highlight the positions of the subjects along the link. 
The obtained root mean square error (RMSE) of the position estimation with CSI and RSSI measurements is $0.57$ m and $0.97$ m respectively (room size
is $6$m). 
To improve accuracy, passive subject position tracking services are implemented by collecting RSSI data from a network of field devices; we further validate the JSON-REST based remote cloud access and its real-time capability.

\subsection{Real-time cloud based localization service} 
The real-time passive localization service collects CQI data from a mesh network of $D=14$ IEEE 802.15.4 field devices (with $V=182$ active links) that are pre-installed inside the monitored area. As depicted in Fig.~\ref{cloud_iot_loc}, the position of the field devices can be arbitrary, however, for 2D imaging purposes, we choose the regular deployment to cover an area of $6\times 3$m. 

The passive subject tracking service is made accessible remotely by an end-user application running on a portable device (inside building number 21). The edge node is installed inside the monitored room (the building labeled as 20) and can be designed to aggregate both RSSI and CSI data. Although not considered in this paper, the edge node is compliant with the OSGi (Open Service Gateway Initiative) framework and can collect/aggregate other data (temperature, luminosity) as well as sensor readings from wearable devices. As introduced before, we use a JSON REST framework to encode the CQI features before posting on the cloud. In particular, the Jax RESTful web services (Jax-RS) are adopted by the cloud section for exposing resources, and use the Jersey-Maven implementation. A list of the available REST resources is illustrated in Fig. \ref{cloud_iot_loc}b.

The remote application opens the socket ``\textbackslash{}api\textbackslash{}localization\textbackslash{}getLatentValues\textquotedblright, to decode the JSON objects that convey information about the task-dependent latent variables, and visualizes the probability values (\ref{eq:ugm-1}) in real-time to obtain a 2D image of the environment. The latent variable values are in-fact encoded as 2D coordinates and are used to map the received probabilities onto a position in the 2D space. 
The position of the tracked subject is also highlighted using a gray-scale code. Selected snapshots\footnote{video records available at www.youtube.com/watch?v=PUBxiJuzvUI, 30 Nov, 2017} of the real-time imaging application and the corresponding ground truth are also depicted in Fig.~\ref{cloud_iot_loc}c.  PHY/MAC-layer limits on CQI data sampling time affect the localization accuracy. In most cases, an observed latency of approx. $100$ms, namely the time delay between a true body movement and its visualization on the 2D image, is associated with peak accelerations of the body running/moving fast or performing rapid/jerky movements. JSON object serialization/deserialization processes also add to the total system latency.
		
\begin{figure}
\includegraphics[trim=0 0 0 0,width=\linewidth]{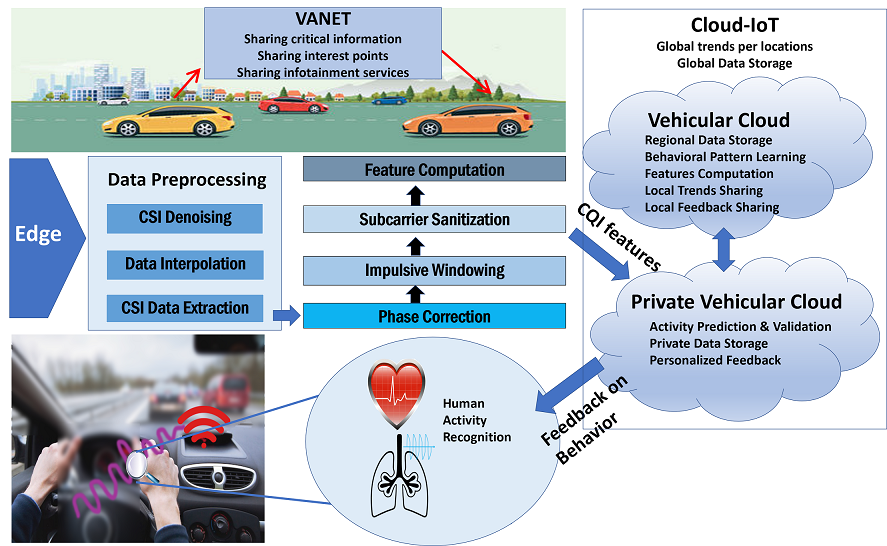}
\caption{Cloud architecture for real-time recognition in vehicles.} 
\label{cloudArchitectCarsFigure}
\end{figure}

\section{Cloud Based Behavior Recognition in Cars\label{sec:Activity-and-Emotion}}
With the advent of half and fully autonomous vehicles, the number of sensors installed in modern cars (GPS, radio devices and embedded computers) are rapidly increasing in order to analyze the users' behavior and to facilitate them with required services.
For emotion or behavior recognition in car the limited capacity of local edge devices is insufficient to handle complex data processing and machine learning steps, moreover, the vehicle-2-vehicle (V2V) communication, behavioral trend learning and data storage can not be efficiently realized locally. 
In order to cater for a wide range of applications from behavior sensing, we thus propose a specific cloud configuration, described in Sect.~\ref{sectionSystemOverview}.

\subsection{System Architecture}\label{sectionSystemOverview}
As depicted in Figure~\ref{cloudArchitectCarsFigure}, the cloud-edge architecture for driver behavior detection inside the car consists of an on-board Gateway (where CQI features are extracted), and a private vehicular cloud profile, including storage and computing for manipulation of features and personalized computing (i.e., of latent variables). 
The cloud system can also interact with a vehicular ad-hoc network (VANET) to provide augmented services. 
We provide a description of the specific design for driver behavior recognition.

\textbf{Edge nodes.}
 Edge nodes, mounted inside the car, process CQI data in a window range of milliseconds to few seconds: features $\mathbf{x}_{e}$ are obtained from PCA (\ref{eq:feat}), but need to be tailored for driver detection, as described in Sect.~\ref{sectionBehaviorRecognition}. The computed features are then sent to the private cloud profile (vehicular cloud).

\textbf{Vehicular cloud profiles.}
A private cloud profile is adopted to perform the detection of individual's behavior and to generate personalized feedback in real-time. 
Such stimuli are inferred by detecting modalities like anomalous body movements or activities. 
Behavior, emotional expressions and feedback preferences adhere to regional preferences. 
Therefore, the behavior detection mechanism can benefit from the utilization of learned features obtained from cars in the same area. 
The private cloud profile facilitates the users to define the data they don't want to share in public, as well as to anonymize the features. 
Vehicular cloud computing~\cite{Gerla} methods can handle such contents while exploiting the increasing mobile storage and processing capacity to privacy offending content on local devices.

\textbf{Applications to VANET.}
Vehicle to Vehicle (V2V) and Vehicle to Infrastructure (V2I) communications platforms~\cite{VANET} can be used to provide early warnings or feedback to nearby moving vehicles about the incidents or other risks. VANETs typically involve spontaneous communities of connected vehicles (of known locations). Connected cars can thus benefit from the proposed driver behavior recognition service as they can share private, anonymized, CQI features, in exchange for information about any risky driving activities in the surroundings.  

\subsection{WiFi Based Behavior Recognition} \label{sectionBehaviorRecognition}
Body movements, gestures and postures reflect emotion and behavior~\cite{Burgoon}.
Inside cars, the driving style might change when the subject is distracted
by internal or external factors. In this section, we focus on the
detection of such distracted behavior. 
We thus detect the changes in body movements by utilizing the non-intrusive, cheap and commercially
off the shelf WiFi field devices described in Sect. \ref{sec:Occupancy-detection-and}.
We capture and analyze the fluctuations of the CSI as induced by erratic body movements. 
In particular, the system detects distracted behavior by monitoring unusual head turns and arm movements during driving situations. 
For instance, if the driver is looking left and right on a straight road, this indicates distraction.
Arm movement indicates whether the subject is performing inappropriate activity like eating, using the smart-phone etc.

\begin{figure*}
\begin{subfigure}{0.33\linewidth}
\includegraphics[trim=0 0 0 0,width=0.99\linewidth] {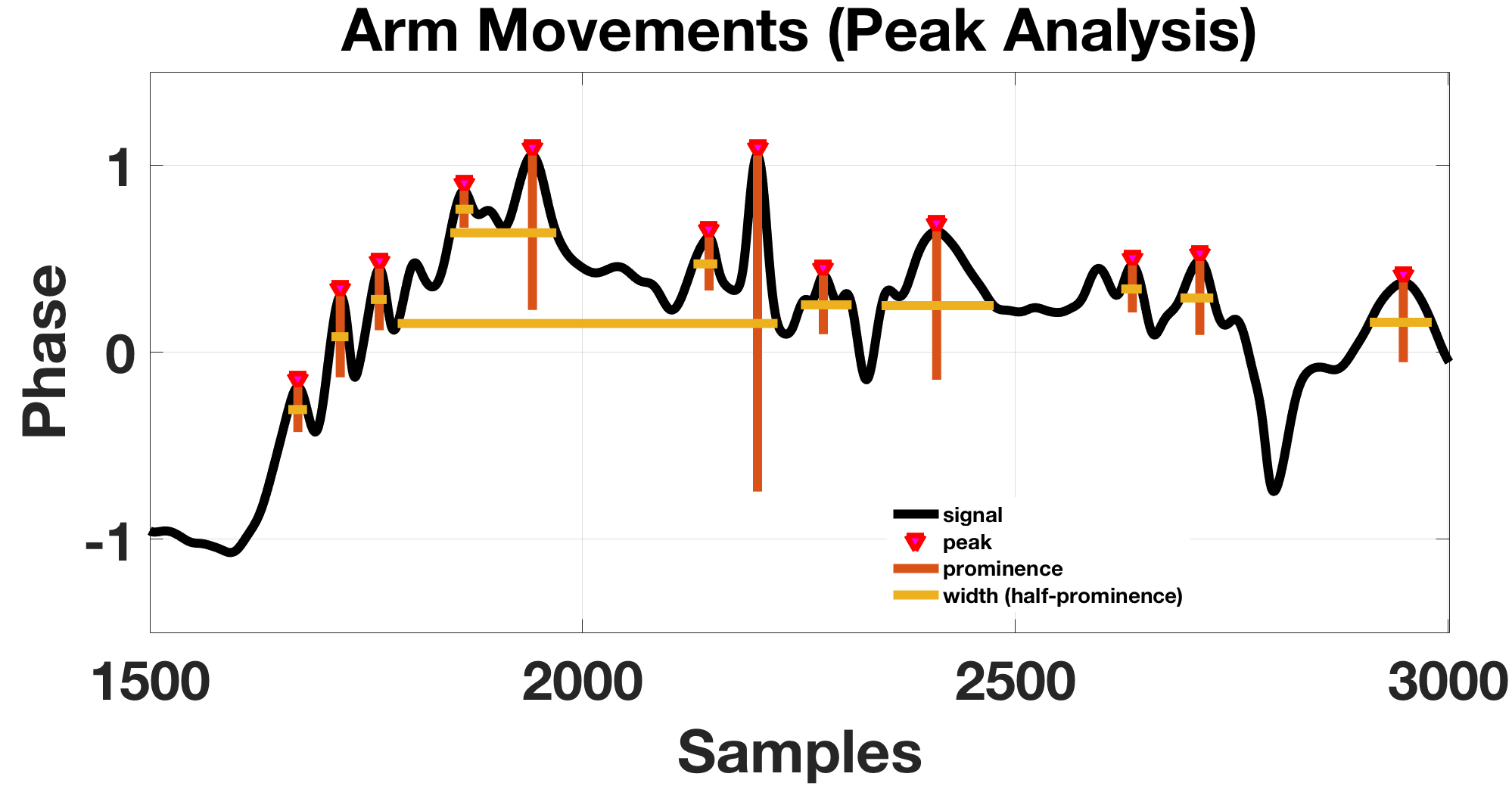}
\caption{Arm Movement.} 
\label{fig:armMovementsFigure}
\end{subfigure}
\begin{subfigure}{0.33\linewidth}
\includegraphics[trim=0 0 0 0,width=0.99\linewidth] {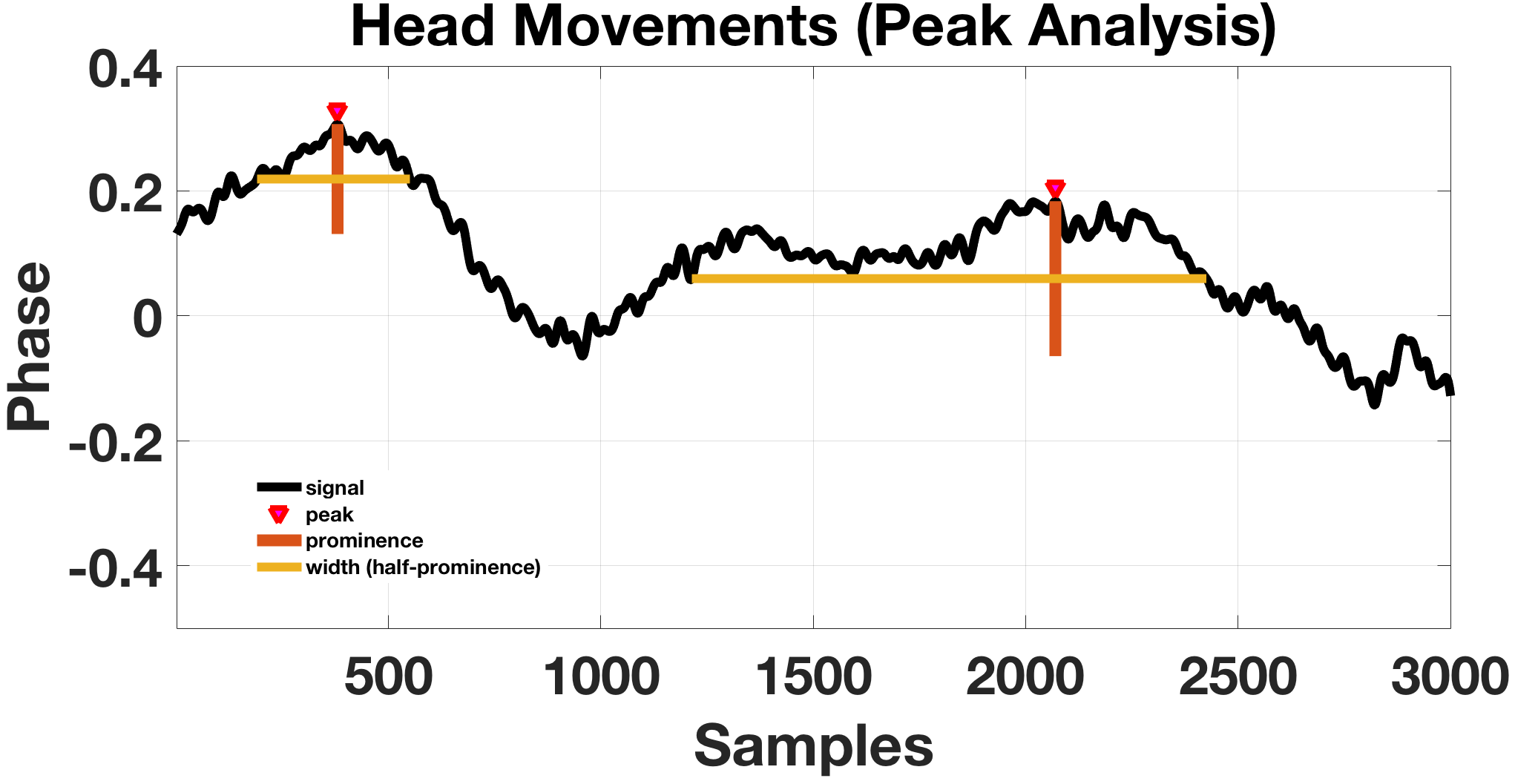}
\caption{head movement.}
\label{fig:headMovementsFigure}
\end{subfigure}
\begin{subfigure}{0.33\linewidth}
\includegraphics[trim=0 0 0 0,width=0.99\linewidth,,height=2.95cm] {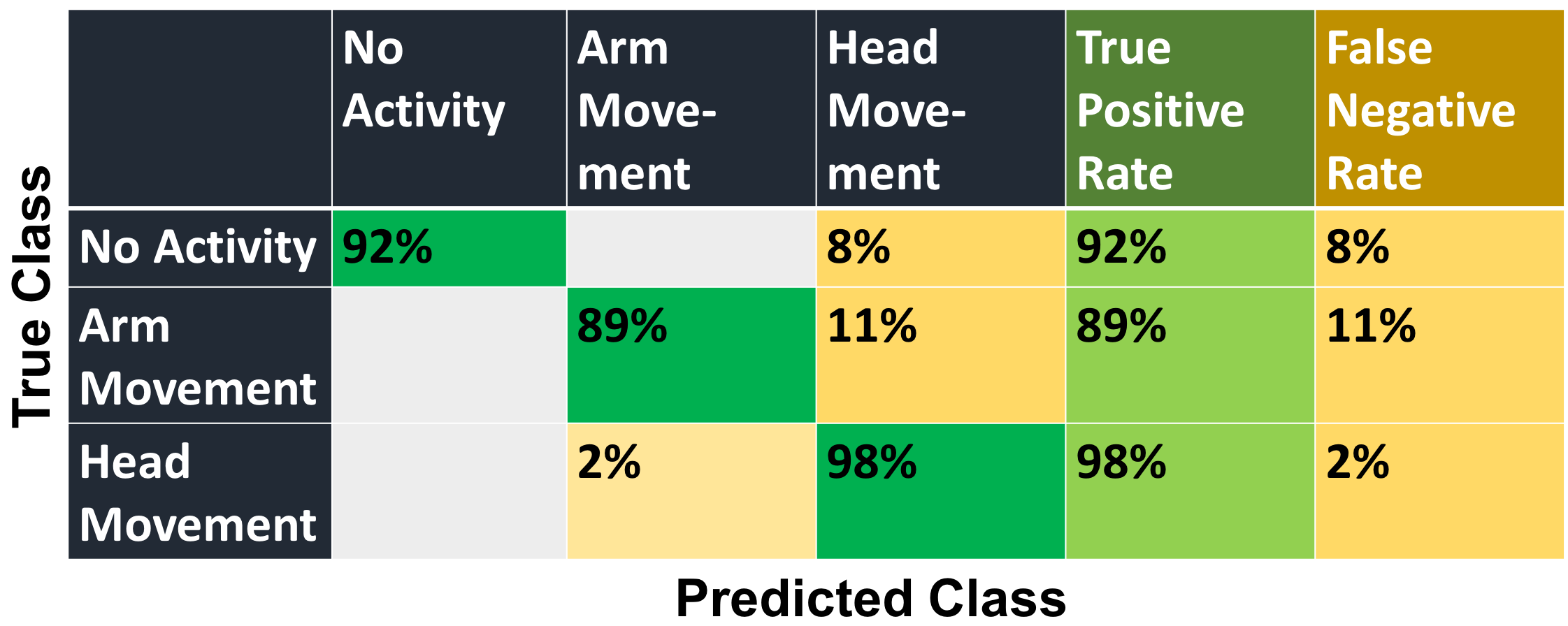}
\caption{Confusion matrix.}
\label{fig:confusionMatrixFigure}
\end{subfigure}
\caption{Peak analysis of CQI features and confusion matrix for head and arm movements.}
\end{figure*}

The recognition consists of several steps. 
First, CSI features are computed in real time obtained from a WiFi - IEEE 802.11n - radio link. 
Then, segmentation identifies the start and end points of changing activities/movements.  
Finally, a feature vector $\mathbf{x}_{e}$ is computed via PCA for CQI and by the analysis of signal peaks from multiple OFDM sub-carriers (more details are given in Sect. \ref{subsec:CQI-feature-computation}).

The cloud profile is responsible to
\begin{enumerate}
\item apply CQI feature manipulation for classifying and predict the activity
labels $z_{i}$ as in equation~(\ref{classification}).
\item generate appropriate feedback conditioned on the behavior detected; 
\item ensure that the system is adaptable and self-learning by collecting feedback
from drivers over time; 
\item test the system to cater for different variables, such as moving vs.
still car, windows open and shut, smooth vs. rough road conditions,
and one vs. multiple passengers. 
\end{enumerate}

\subsection{Study design}
We collected data of different driver behaviors by conducting a human study with 40 participants at BMW Group Research, New Technologies and Innovation center, Germany. 
We have acquired 3 separate classes of data:
one for calibration purpose, to obtain training data-sets $\mathbf{\overline{s}}_{d}(z_{i}=k|\tau_{i})$, a second one, for distracted behavior driving and a third one for relaxed driving.

The subjects drove in a simulated road environment, and the
distractions are induced by unknown, random and repetitive triggers
which resulted head turns, gestures and arm movement. 
In particular, we used the BMW Mini Cooper and Augmented Reality simulation video tool. 
For each subject, the driving time is around 15 minutes. 
The driver might move his head or arms in different directions in response to specific continuous triggers acting as random distractions. 
The driving simulation was identical for all subjects while the trigger sequence was random. 
More than 80\% of the subjects reported that they felt distracted and annoyed during the drive.

\textbf{Hardware setup.}
\label{subsectionHardwarePrototype} Our hardware prototype consists of a single transmitter and receiver. The WiFi chipset is installed inside a portable laptop for capturing CSI data with the modified Linux driver (see Sect. \ref{sec:Occupancy-detection-and}). Similarly as in the previous indoor localization tests, ground truth videos are recorded as well as timestamps for each packet to ensure correct labeling\footnote{using the ELAN http://tla.mpi.nl/tools/tla-tools/elan/ open source tool}.

\textbf{CQI feature computation and classification.}\label{subsec:CQI-feature-computation}
To achieve accurate detection results, we pre-manipulate the CSI data
by using phase-based denoising and change detection tools \cite{change}, to separate the human movements and to distinguish between different activities. The CQI feature vector $\mathbf{x}_{e}$
contains the principal components (equation~(\ref{eq:feat})). 
However, compared with passive localization, behavior detection requires more information about the observed CSI patterns. Therefore, additional features are collected from i) a peak analysis of the segmented CSI data obtained after PCA and ii) analysis of the deviations of the CSI phase information. These added features are crucial to distinguish between head and arm movements.

\begin{enumerate}
\item \textit{CSI signal peak analysis.} Peak analysis extracts the following features (see the examples in Fig. \ref{fig:armMovementsFigure} for arm, and Fig. \ref{fig:headMovementsFigure} for head movements): i) the number of peaks (including inverted peaks), ii) peak (and inverted) width. 
We observe (Fig.~\ref{fig:armMovementsFigure} and~\ref{fig:headMovementsFigure}) that peak characteristics of arm movements are different from head movements. 
For example, the number of peaks corresponding to a single arm movement is greater than for a head movement, while peak width is larger in head movements. 
In-fact, arm movements typically cause additional jerky/voluntary body motions and have thus a larger impact on CSI footprints.

\item \textit{OFDM subcarrier analysis.} The analysis of OFDM subcarriers
gives further information about driver movements inside the car. 
In particular, the fluctuation (or standard deviation) of the CSI phase difference between consecutive subcarriers, namely $\sigma_{t,\ell_{j}}=\angle{s_{f_{i},\ell_{j},t}}-\angle{s_{f_{i-1},\ell_{j},t}}$, is another property that can be used to distinguish between head and arm movements. 
Therefore, the feature vector $\mathbf{x}_{e}$ also contains: i) the average deviation $\mathrm{E}_{t}(\sigma_{t,\ell_{j}})$ w.r.t. the considered time window, and ii) the minimum deviation, i.e., $\mathrm{min}_{t}(\sigma_{t,\ell_{j}})$. In head movements, we noticed that the variance
of phase difference between the subcarriers is widespread, while it
significantly drops in the case of the larger arm movements.
\end{enumerate}


The edge layer in Figure~\ref{cloudArchitectCarsFigure} shows the
steps of our behavior detection system based on WiFi devices. At first,
the system interpolates the data to cater for missing packets. 
The phase in CSI data is corrected to make it usable for analysis. We then employ noise
removal in our denoising step. 
Then, we mark the boundaries of activities via change detection tools \cite{change} and we apply PCA on segmented data. The peak and phase difference analysis is performed to compute the additional features.
The whole feature set is finally pushed to the cloud profile for classification
using equation~(\ref{classification}).

Impulsive windowing and critical choice of features
make the classification model learning a simple process. Our feature vector is composed of features from PCA as well as peak and subcarrier analysis. We then perform classification by first separating data into test and
train samples. We use K-Nearest neighbour, with $6$ neighbours for
learning. Therefore, the mixture component $\mathrm{G}^{k}(\mathbf{x}_{e}|\tau_{i})$
in equation~(\ref{eq:ugm-2}) corresponds to the similarity (euclidean distance) between the observed features $\mathbf{x}_{e}$ and the expected ones, considering the latent process value $z_{i}=k$. 
The train and test sample data is randomly distributed
from different subjects to include all possible movements for
training. Performing 10-fold cross-validation and separated
test samples gives us overall accuracy of 94.5\% as depicted in the confusion matrix shown in Fig.~\ref{fig:confusionMatrixFigure} for
separating arm movements from head movements and no activity. The training speed of the system is approx. $1200$ samples/sec. while the training time is approx. $1.58$ sec. Overall, the false negative rate for the arm movements is the largest, being 11\%: this is due to a wider range and intensity of arm movements compared with head movements.

\section{Concluding remarks and open problems}
We presented our cloud-IoT platform for passive radio sensing. 
In particular, we highlighted the main challenges and the specifications of the cloud architecture, the corresponding models and resources. 
Real time cloud-edge analytics of different CQI data types (i.e., RSSI and CSI) were investigated by focusing on data acquisition, processing, low-dimensional CQI feature representation, extraction and manipulation. The proposed cloud-IoT structure interacts with remote end-user applications to support different passive radio sensing tasks. 
It reconfigures CQI data collection and can be tailored for integration with existing sensing cloud platforms and semantic data models. Validation of cloud functions is based on two case studies, focusing on indoor localization in a smart laboratory space and in-car driver behavior recognition using ZigBee and WiFi radio technologies, respectively.     

The platform can be easily adapted to implement different sensing tasks, as well as to foster application orchestration. 
New tasks might require additional application-specific modules to adapt the CQI feature manipulation stage. On the other hand, all tasks can operate on shared CQI data models, interfaces and cloud-edge tools.
Future work will focus on advanced techniques for sensor fusion and integration of emerging high frequency technologies (from 60GHz to 150GHz). 


\end{document}